\documentclass[aps,showpacs,amssymb,eqsecnum]{revtex4}
\usepackage{graphicx}
\begin{document}
\title
{
Damped finite-time-singularity
driven by noise
}
\author{Hans C. Fogedby}
\email{fogedby@phys.au.dk}
\affiliation { Institute of Physics and
Astronomy,
University of Aarhus, DK-8000, Aarhus C, Denmark\\
and\\
NORDITA, Blegdamsvej 17, DK-2100, Copenhagen {\O}, Denmark
}
\date{\today}
\begin{abstract}
We consider the combined influence of linear damping and noise on
a dynamical finite-time-singularity model for a single degree of
freedom. We find that the noise effectively resolves the
finite-time-singularity and replaces it by a first-passage-time or
absorbing state distribution with a peak at the singularity and a
long time tail. The damping introduces a characteristic cross-over
time. In the early time regime the probability distribution and
first-passage-time distribution show a power law behavior with
scaling exponent depending on the ratio of the non linear coupling
strength to the noise strength. In the late time regime the
behavior is controlled by the damping. The study might be of
relevance in the context of hydrodynamics on a nanometer scale, in
material physics, and in biophysics.
\end{abstract}
\maketitle
\section{\label{secintro}Introduction}
The influence of noise on the behavior of nonlinear dynamical
system is a recurrent theme in modern statistical physics
\cite{Freidlin98}. In a particular class of systems the nonlinear
character gives rise to finite-time-singularities, that is
solutions which cease to be valid beyond a particular finite time
span. One encounters finite-time-singularities in stellar
structure, turbulent flow, and bacterial growth
\cite{Kerr99,Brenner97,Brenner98}. The phenomenon is also seen in
Euler flows and in free-surface-flows \cite{Cohen99,Eggers97}.
Finally, finite-time-singularities are encountered in modelling in
econophysics, geophysics, and material physics
\cite{Sornette02,Ide01,Sornette01b,Gluzman01a,Johansen02}.

In the context of hydrodynamical flow on a nanoscale
\cite{Eggers02}, where microscopic degrees of freedom come into
play, it is a relevant issue how noise influences the
hydrodynamical behavior near a finite-time-singularity. Leaving
aside the issue of the detailed reduction of the hydrodynamical
equations to a nanoscale and the influence of noise on this scale
to further study, we assume in the present context that a single
variable or ``reaction coordinate'' effectively captures the
interplay between the singularity and the noise.

In a recent paper \cite{Fogedby02d} we investigated a simple
generic model system with  one degree of freedom governed by a
nonlinear Langevin equation driven by Gaussian white noise,
\begin{eqnarray}
\frac{dx}{dt} = - \frac{\lambda}{2|x|^{1+\mu}} + \eta ~,~~~
\langle\eta\eta\rangle(t) = \Delta\delta(t). \label{lan1}
\end{eqnarray}
The model is characterized by the coupling parameter $\lambda$,
determining the amplitude of the singular term, the index $\mu\geq
0$, characterizing  the nature of the singularity, and the noise
parameter $\Delta$, determining the strength of the noise
correlations. Specifically, in the case of a thermal environment
at temperature $T$ the noise strength $\Delta\propto T$.

In the absence of noise this model exhibits a
finite-time-singularity at a time $t_0$, where the variable $x$
vanishes with a power law behavior determined by $\mu$. When noise
is added  the finite-time-singularity event at $t_0$ becomes a
statistical event and is conveniently characterized by a
first-passage-time distribution $W(t)$ \cite{Redner01}. For
vanishing noise we have $W(t)=\delta(t-t_0)$, restating the
presence of the finite-time-singularity. In the presence of noise
$W(t)$ develops a peak about $t=t_0$, vanishes at short times, and
acquires a long time tail.

The model in Eq. (\ref{lan1}) has also been studied in the context
of persistence distributions related to the nonequilibrium
critical dynamics of the two-dimensional XY model \cite{Bray00}
and in the context of non-Gaussian Markov processes
\cite{Farago00}. Finally, regularized for small $x$, the model
enters in connection with an analysis of long-range correlated
stationary processes \cite{Lillo02}.

From our analysis in ref. \cite{Fogedby02d} it followed that for
$\mu=0$, the logarithmic case, the distribution at long times is
given by the power law behavior
\begin{eqnarray}
W(t)\sim
t^{-\alpha}~~,~~~\alpha=\frac{3}{2}+\frac{\lambda}{2\Delta}.
\label{pow}
\end{eqnarray}
For vanishing nonlinearity, i.e., $\lambda=0$, the
finite-time-singularity is absent and the Langevin equation
(\ref{lan1}) describes a simple random walk of the reaction
coordinate, yielding the well-known exponent $\alpha=3/2$
\cite{Redner01,Stratonovich63,Risken89}. In the nonlinear case
with a finite-time-singularity the exponent attains a
non-universal correction, depending on the ratio of the nonlinear
strength to the strength of the noise; for a thermal environment
the correction is proportional to $1/T$. In the generic case for
$\mu>0$ we found that the fall-off is slower and that the
correction to the random walk result is given by a stretched
exponential
\begin{eqnarray}
W(t)\sim t^{-3/2}\exp[-A(t^{-\mu/(2+\mu)}-1)], \label{stret}
\end{eqnarray}
where $A\rightarrow\lambda/\Delta\mu$ for $\mu\rightarrow 0$.

In our studies so far we have ignored damping. It is, however,
clear that in realistic physical situations friction or damping
must enter on the same footing as the noise. This follows from the
Einstein relation or more generally from the
fluctuation-dissipation theorem relating the damping to the noise.
In the present paper we attempt to amend this situation and thus
proceed to extend the analysis in ref. \cite{Fogedby02d} to the
case of linear damping. We shall here only consider the
logarithmic case for $\mu=0$.

For this purpose we consider the following model for one degree of
freedom:
\begin{eqnarray}
\frac{dx}{dt} = -\gamma x - \frac{\lambda}{2|x|} + \eta ~,~~~
\langle\eta\eta\rangle = \Delta\delta(t). \label{lan2}
\end{eqnarray}
In addition to the coupling parameter $\lambda$, and the noise
parameter $\Delta$, this model is also characterized by the
damping constant $\gamma$. Assuming for convenience a
dimensionless variable $x$ the coupling and the noise strengths
$\lambda$ and $\Delta$ have the dimension $1/\text{time}$. The
ratios $\lambda/\Delta$ and $\gamma/\Delta$ are thus dimensionless
parameters characterizing the behavior of the system.

It follows from our analysis below that the damping constant sets
an inverse time scale $1/\gamma$. At intermediate time scales for
$\gamma t\ll 1$ the distribution exhibits the same power law
behavior as in the undamped case given by Eq. (\ref{pow}). At long
times for $\gamma t\gg 1$ the distribution falls off exponentially
with time constant $1/\gamma(1+\lambda/\Delta)$, i.e.,
\begin{eqnarray}
W(t)\propto\exp[-\gamma(1+\lambda/\Delta)t].
\end{eqnarray}

The paper is organized in the following manner. In Section
\ref{secmodel} we introduce the finite-time-singularity model with
linear damping and discuss its properties. In Section \ref{secwkb}
we review the  weak noise WKB phase space approach to the
Fokker-Planck equation, apply it to the finite-time-singularity
problem with damping, discuss the associated dynamical phase space
problem and the long time properties of the distributions. In
Section \ref{secfokker} we derive an exact solution of the
Fokker-Planck equation and present an expression for the
first-passage-time distribution. In Section \ref{secsum} we
present a summary and a conclusion. In the present treatment we
draw heavily on the analysis in ref. \cite{Fogedby02d}. In
Appendix \ref{app1} aspects of the exact solution are discussed in
more detail; in Appendix \ref{app2} we consider the weak noise
limit of the exact solution.
\section{\label{secmodel}Model}
\begin{figure}
\includegraphics[width=.8\hsize]{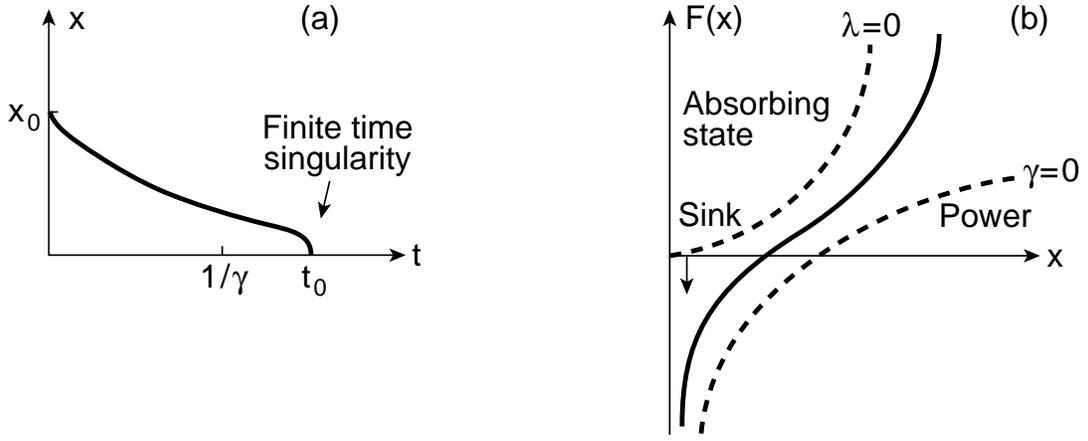}
\caption{ In a) we show the time evolution of the single degree of
freedom $x$. At times shorter than the cross-over time $1/\gamma$
the variable $x$ falls off exponentially. At times beyond
$1/\gamma$ the variable $x$ reaches the absorbing state $x=0$ at a
finite time $t_0$. In b) we depict the free energy $F(x)$ driving
the equation. For $\lambda=0$ the free energy forms a confining
harmonic well, for $\gamma=0$ we have the absorbing state case
discussed in ref. \cite{Fogedby02d}. In the general case the
absorbing state $x=0$ corresponds to the sink in $F(x)$.}
\label{fig1}
\end{figure}

In terms of a free energy or potential $F$ we can express Eq.
(\ref{lan2}) in the form
\begin{eqnarray}
\frac{dx}{dt} = - \frac{1}{2}\frac{dF}{dx}+ \eta(t),\label{lan3}
\end{eqnarray}
where $F$ has the form
\begin{eqnarray}
F= \gamma x^2 +\lambda\ln|x|.
\end{eqnarray}
The free energy has a logarithmic sink and drives $x$ to the
absorbing state $x=0$. For large $x$ the free energy has the form
of a harmonic well potential confining the motion. In
Fig.~\ref{fig1} we have depicted the free energy in the various
cases.
\subsection{The noiseless case}
In the case of vanishing noise Eq. (\ref{lan3}) is readily solved.
We obtain
\begin{eqnarray}
x(t)=\left[\frac{\lambda}{2\gamma}\right]^{1/2} \left[e^{2\gamma
(t_0-t)}-1\right]^{1/2}, \label{sol}
\end{eqnarray}
with a finite-time-singularity at
\begin{eqnarray}
t_0 =
\frac{1}{2\gamma}\log\left|1+\frac{2\gamma}{\lambda}x_0^2\right|.
\end{eqnarray}
The initial value of $x$ is $x_0$ at time $t=0$. In the presence
of damping $x$ initially falls off exponentially due to the
confining harmonic potential with a time constant $1/\gamma$. For
times beyond $1/\gamma$ the nonlinear term takes over and drives
$x$ to zero at time $t_0$, i.e.,  $x$ falls into the sink in $F$.
In Fig.~\ref{fig1} we have shown the noiseless solution $x(t)$.
%

\subsection{The noisy case}
Summarizing the discussion in ref. \cite{Fogedby02d}, the
stochastic aspects of the finite-time-singularity in the presence
of noise are analyzed by focusing on the time dependent
probability distribution $P(x,t)$ and the derived
first-passage-time or absorbing state probability distribution
$W(t)$. The distribution $P(x,t)$ is defined according to
\cite{vanKampen92,Risken89} $P(y,t)=\langle\delta(y-x(t))\rangle$
where $x$ is a stochastic solution of Eq. (\ref{lan3}) and
$\langle\cdots\rangle$ indicates an average over the noise $\eta$
driving $x$. In the absence of noise $P(y,t)=\delta(y-x(t))$,
where $x$ is the deterministic solution given by Eq. (\ref{sol})
and  depicted in Fig~\ref{fig1}. At time $t=0$ the variable $x$
evolves from the initial condition $x_0$, implying the boundary
condition $P(x,0)=\delta(x-x_0)$.

At short times  $x$  is close to $x_0$  and the singular term and
the damping term are  not yet operational. In this regime we
obtain ordinary random walk with the Gaussian distribution,
$P(x,t)=(2\pi\Delta t)^{-1/2} \exp\left[-(x-x_0)^2/2\Delta
t\right]$. At a time scale given by $1/\gamma$ the damping drives
$x$ towards a stationary distribution given by
$P\propto\exp({-F/\Delta)}$. However, at longer times beyond the
scale $1/\gamma$, the barrier $\lambda/2x$ comes into play
preventing $x$ from crossing the absorbing state $x=0$. This is,
however, a random event which can occur at an arbitrary time
instant, i.e., the finite-time-singularity at $t_0$ in the
deterministic case is effectively resolved in the noisy case. For
not too large noise strength the distribution  is peaked about the
noiseless solution and vanishes for $x\rightarrow 0$,
corresponding to the absorbing state, implying the boundary
condition
\begin{eqnarray}
P(0,t)=0~. \label{bou2}
\end{eqnarray}
In order to  model a possible experimental situation the
first-passage-time or absorbing state distribution $W(t)$ is of
more direct interest \cite{vanKampen92,Gardiner97}.

Since $P(0,t)=0$ for all $t$  due to the absorbing state, the
probability that $x$ is not reaching $x=0$ in time $t$ is thus
given by $\int_0^\infty P(x,t)dx$, implying  that the probability
$-dW$ that  $x$ does reach $x=0$ in time $t$ is
$-dW=-\int_0^\infty dx dt (dP/dt)$, yielding the absorbing state
distribution $W(t) =-\int_0^\infty dx~\partial P(x,t)/\partial t$
\cite{Risken89}. In the absence of noise $P(x,t)=\delta(x-x(t))$
and $W(t)=\delta(t-t_0)$, in accordance with the finite time
singularity at $t=t_0$. For weak noise $W(t)$ peaks about $t_0$
with vanishing tails for small $t$ and large $t$.

The distribution $P(x,t)$ satisfies the Fokker-Planck equation
\cite{vanKampen92,Gardiner97}
\begin{eqnarray}
\frac{\partial P}{\partial t} =
\frac{1}{2}\frac{\partial}{\partial x}
\left[\frac{dF}{dx}P+\Delta\frac{\partial P} {\partial x}
\right]~, \label{fokker}
\end{eqnarray}
in the present case subject to the boundary conditions
$P(x,0)=\delta(x-x_0)$ and $P(0,t)=0$. The Fokker-Planck equation
has the form of a conservation law $\partial P/\partial t +
\partial J/\partial x=0$, defining the probability current
$J=(1/2)(dF/dx)P-(1/2)\Delta\partial P/\partial x$ and we obtain
for $W(t)$ the expression
\begin{eqnarray}
W(t)=\frac{1}{2}\left[\frac{dF}{dx}P+\Delta\frac{\partial
P}{\partial x}\right]_{x=0}, \label{absd2}
\end{eqnarray}
to be used in our further analysis. Note that there is a sign
error in Eq. (3.8) in ref. \cite{Fogedby02d}.
\section{\label{secwkb}Weak noise approach}
In this section we apply a weak noise canonical phase space
approach to the damped finite-time-singularity model and infer the
general long time behavior. The treatment follows closely the
analysis in ref. \cite{Fogedby02d}.
\subsection{The phase space method}
From a structural point of view the Fokker-Planck equation
(\ref{fokker}) has the form of an imaginary-time Schr\"{o}dinger
equation $\Delta\partial P/\partial t=HP$, driven by the
Hamiltonian or Liouvillian $H$. The noise strength $\Delta$ plays
the role of an effective Planck constant and $P$ corresponds to
the wavefunction. Drawing on this parallel we have in recent work
in the context of the Kardar-Parisi-Zhang equation for a growing
interface elaborated on a weak noise nonperturbative WKB phase
space approach to a generic Fokker-Planck equation for extended
system \cite{Fogedby99a,Fogedby99b,Fogedby02a}. In the case of a
single degree of freedom this method amounts to the eikonal
approximation \cite{Risken89,Roy93,Gardiner97}, see also
\cite{Graham73,Graham89}. For systems with many degrees of freedom
the method has for example been expounded in \cite{Falkovich96},
based on the functional formulation of the Langevin equation
\cite{Martin73,Janssen76}. In the present formulation
\cite{Fogedby99a,Fogedby99b,Fogedby02a} the emphasis is on the
canonical phase space analysis and the use of dynamical system
theory \cite{Strogatz94,Ott93}.

The weak noise WKB approximation corresponds to the ansatz
$P\propto\exp[-S/\Delta]$. The weight function or action $S$ then
to leading asymptotic order in $\Delta$ satisfies a
Hamilton-Jacobi equation $\partial S/\partial t+H=0$ which in turn
implies a {\em principle of least action} and Hamiltonian
equations of motion \cite{Landau59b,Goldstein80}. In the present
context the Hamiltonian takes the form
\begin{eqnarray}
H=\frac{1}{2}p(p-\frac{\lambda}{x}-2\gamma x), \label{ham1}
\end{eqnarray}
yielding the Hamilton equations of motion
\begin{eqnarray}
&&\frac{dx}{dt}= -\gamma x - \frac{\lambda}{2x}+p, \label{eqx}
\\
&&\frac{dp}{dt}=\gamma p -\frac{1}{2}\frac{\lambda}{x^2}p.
\label{eqp}
\end{eqnarray}
These equations replace the Langevin equation (\ref{lan2}) with
the noise $\eta$ represented  by the momentum $p=\partial
S/\partial x$, conjugate to $x$. The equations (\ref{eqx}) and
(\ref{eqp}) determine orbits in a canonical phase space spanned by
$x$ and $p$. Since the system is conserved the orbits lie on the
constant energy manifold(s) given by $E=H$. The action associated
with an orbit from $x_0$ to $x$ in time $t$ has the form
\begin{eqnarray}
S(x_0\rightarrow x,t)=\int_0^t dt\left[p\frac{dx}{dt}-H\right].
\label{action}
\end{eqnarray}
According to the ansatz the probability distribution is then given
by
\begin{eqnarray}
P(x,t)= P(x_0\rightarrow x,t)
\propto\exp\left[-\frac{S(x_0\rightarrow x,t)}{\Delta}\right]~.
\label{dis2}
\end{eqnarray}
\subsection{Long time orbits}
\begin{figure}
\includegraphics[width=.8\hsize]{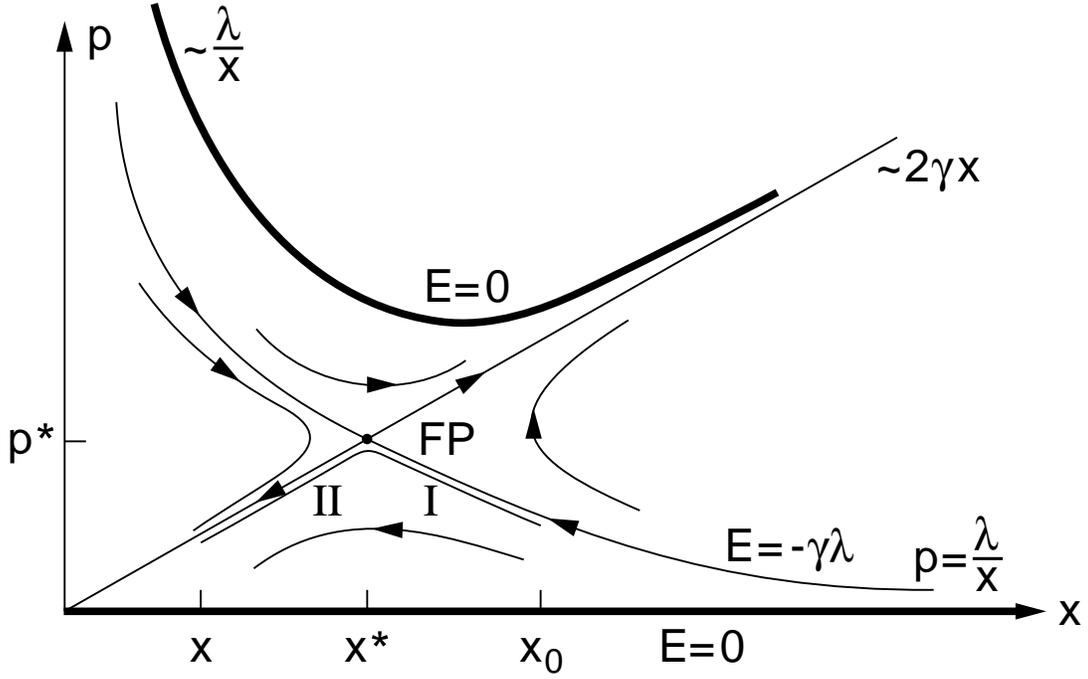}
\caption{We show the topology of phase space.  The bold lines
indicate the zero energy sub-manifolds. The invariant heteroclinic
orbits $p=2\gamma x$ and $p=\lambda/x$ passing through the saddle
point FP at $(x^\ast,p^\ast)=
[(\lambda/2\gamma)~{1/2},(2\gamma\lambda)^{1/2}]$ have energy
$-\gamma\lambda$. The arrows indicate the direction of the flow.
The long time orbit from $x_0$ to $x$ passes close to the fixed
point. The part of the orbit following the invariant manifold
$p=\lambda/x$ and entering in our long time estimate is denoted I;
the part of the orbit close to the $p=2\gamma x$ manifold is
denoted II.} \label{fig2}
\end{figure}

The zero-energy manifolds delimiting the phase space orbits follow
from Eq. (\ref{ham1}) and are given $p=0$ and $p=2\gamma x
+\lambda/2x$. The $p=0$ sub-manifold corresponds to the noiseless
or deterministic case discussed above. The $p=\lambda/x+2\gamma x$
sub-manifold corresponds to the noisy case. By insertion in Eq.
(\ref{eqp}) we obtain $dx/dt=\gamma x + \lambda/2x$, i.e., the
motion on the noisy sub-manifold is time reversed of the motion on
the noiseless sub-manifold. The orbit structure in phase space is
moreover controlled by the hyperbolic fixed point at
$(x^\ast,p^\ast)=[(\lambda/2\gamma)^{1/2},(2\gamma\lambda)^{1/2}]$.
The heteroclinic orbits passing through the fixed point are given
by $p=\lambda/x$ and $p=2\gamma x$ and the energy of the invariant
manifold is $E^\ast=-\gamma\lambda$. In Fig~\ref{fig2} we have
depicted the phase space with the zero-energy manifolds, the fixed
point, the heteroclinic orbits and some characteristic orbits.

The long time behavior of the distribution is determined by an
orbit from $x_0$ to $x$ traversed in time $t$. In the long time
limit this orbit must pass close to the hyperbolic fixed point.
Note that in the limit $\gamma\rightarrow 0$ the fixed point
migrates to infinity in the $x$ direction and the long time orbits
approach the zero energy sub-manifolds which thus determine the
asymptotic properties as discussed in ref. \cite{Fogedby02d}.

Independent of whether the initial value $x_0$ is greater or
smaller than the fixed point value $x^\ast$, the long time orbit
follows the invariant $p=\lambda/x$ manifold towards the fixed
point. At the fixed point the orbit slows down and then speeds up
again as the orbit follows the other invariant manifold $p=2\gamma
x$ towards the endpoint $x$ reached in time $t$. This behavior is
also depicted in Fig~\ref{fig2}.

This scenario allows a simple analysis of the long time behavior
of the distribution $P(x,t)$ and the first-passage-distribution
$W(t)$. Close to the invariant manifolds with energy
$E^\ast=-\gamma\lambda$ the action associated with an orbit from
$x_0$ to $x$ follows from Eq. (\ref{action}) and is given by
\begin{eqnarray}
S=-E^\ast t +\int_{x_0}^{x^\ast}dx\frac{\lambda}{x} +
\int_{x^\ast}^x dx 2\gamma x,
\end{eqnarray}
or, denoting the relevant manifolds by a subscripts, see
Fig.~\ref{fig2},
\begin{eqnarray}
S=\lambda\gamma t +
\lambda\log\left|\frac{x^\ast}{x_0}\right|_{\text{I}}+\gamma(x^2-x^{\ast
2})_{\text{II}}. \label{action2}
\end{eqnarray}
At long times  we only have to consider the contribution from the
orbit leading up to the fixed point. Inserting the manifold
condition $p=\lambda/x$ in the equation of motion (\ref{eqx}) we
thus obtain $dx/dt=-\gamma x + \lambda/2x$ with solution
\begin{eqnarray}
x(t)^2 =x_0^2e^{-2\gamma t} + x^{\ast 2}(1-e^{-2\gamma t}).
\label{sol1}
\end{eqnarray}
\subsection{Discussion}
It follows from Eq. (\ref{sol1}) that the damping $\gamma$ sets an
inverse time scale delimiting two kinds of characteristic
behavior. First, for $t\rightarrow\infty$ the orbit approaches the
fixed point $x^\ast$. For $\gamma t\gg 1$ we have $x^2=x^{\ast
2}(1-\exp(-2\gamma t))$ and $x$ approaches the fixed point in an
exponential fashion. On the other hand, in the intermediate time
region for $\gamma t\ll 1$ and for $\gamma t\gg x_0^2$ and
$\lambda t\gg x_0^2$ we obtain $x^2=x_0^2+2t\gamma
x^\ast=x_0^2+\lambda t\sim \lambda t$.

By insertion in the expression (\ref{action2}) for the action we
then obtain in the late time regime for $\gamma t\gg 1$
\begin{eqnarray}
S(t)\sim\lambda\gamma t -\frac{\lambda}{2}e^{-2\gamma t},
\end{eqnarray}
yielding the distribution and ensuing first-passage-time
distribution
\begin{eqnarray}
P(t)\propto W(t)\propto \exp(-\lambda\gamma t/\Delta).
\end{eqnarray}
Likewise, we have in the intermediate time regime $\gamma t\ll 1$
\begin{eqnarray}
S(t)\sim\lambda\gamma t +\frac{\lambda}{2}\log|t|,
\end{eqnarray}
giving rise to the distribution and first-passage-time
distribution
\begin{eqnarray}
P(t)\propto W(t)\propto |t|^{-\frac{\lambda}{2\Delta}}.
\end{eqnarray}

These results hold in the weak noise limit. We note that at long
times $W(t)$ falls of exponentially with a time constant given by
$\Delta/\lambda\gamma$. In the intermediate time regime $W(t)$
exhibits a power law behavior with exponent $-\lambda/2\Delta$,
independent of $1/\gamma$ defining the cross-over time. These
results will also be recovered from the exact solution discussed
in the next sections.
\section{\label{secfokker}Exact solution}
In this section we return to the Fokker-Planck equation
(\ref{fokker}) and present an exact solution. This solution is an
extension of the solution presented in ref. \cite{Fogedby02d} and
the analysis proceeds in much the same way. Details are discussed
in Appendices \ref{app1} and \ref{app2}.
\subsection{Quantum particle in a harmonic potential with
centrifugal barrier}
The Fokker Planck equation has the form
\begin{eqnarray}
\frac{\partial P}{\partial t} = \frac{\Delta}{2}\frac{\partial^2
P}{\partial x^2} + \left(\gamma
x+\frac{\lambda}{2x}\right)\frac{\partial P}{\partial x} +
\left(\gamma -\frac{\lambda}{2x^2}\right)P. \label{fokker2}
\end{eqnarray}
Eliminating the first order term by means of the gauge
transformation
\begin{eqnarray}
\exp(h)=|x|^{-\lambda/2\Delta}e^{-\gamma x^2/2\Delta},
\end{eqnarray}
we can express the equation in the form
\begin{eqnarray}
-\Delta\frac{\partial}{\partial t}\left[\exp(-h)P\right] =
H\left[\exp(-h)P\right], \label{fokker3}
\end{eqnarray}
where the Hamiltonian $H$ driving $P$ is given by
\begin{eqnarray}
H=-\frac{1}{2}\Delta^2\frac{\partial^2}{\partial x^2} +
\frac{\lambda^2}{8}\left[1+\frac{2\Delta}{\lambda}\right]
\frac{1}{x^2} +
\frac{\Delta\gamma}{2}\left(\frac{\lambda}{\Delta}-1\right)
+\frac{\gamma^2}{2}x^2. \label{ham2}
\end{eqnarray}
This Hamiltonian describes the motion of a unit mass quantum
particle in one dimension in a harmonic potential subject to a
centrifugal barrier of strength $(\lambda^2/8)(1+2\Delta/\lambda)$
at the origin; $\Delta$ plays the role of an effective Planck
constant. Note that in  Eq. (6.4) in ref. \cite{Fogedby02d} the
factor $\Delta/2$ should read $1/\Delta$.

For $\lambda=0$ and $\gamma=0$ both the barrier and the confining
potential are absent; the spectrum of $H$ forms a band and the
particle can move over the whole axis. This case corresponds to
ordinary random walk \cite{Risken89}. Incorporating the absorbing
state condition in Eq. (\ref{bou2}) by means of the method of
mirrors we obtain the results presented in ref. \cite{Fogedby02d},
i.e.,
\begin{eqnarray}
P(x,t)=(2\pi\Delta
t)^{-1/2}\left[\exp\left[-\frac{(x-x_0)^2}{2\Delta t}\right] -
\exp\left[-\frac{(x+x_0)^2}{2\Delta t}\right]\right],
\end{eqnarray}
in the half space $x\geq 0$, and for the absorbing state
distribution
\begin{eqnarray}
W(t)=\left(\frac{2}{\pi}\right)^{1/2} x_0\exp(-x_0^2/2\Delta
t)(\Delta t)^{-3/2}.
\end{eqnarray}

For $\lambda\neq 0$ and $\gamma=0$ the particle cannot cross the
barrier and is confined to either half space; this corresponds to
the case of a finite-time-singularity subject to noise and an
absorbing state at $x=0$ and was discussed in detail in ref.
\cite{Fogedby02d}; for reference we give the obtained results
below. Note that $x$ and $x_0$ should be interchanged in Eq. (6.5)
and that a factor $\Delta$ is missing in Eq. (6.6) in ref.
\cite{Fogedby02d}.
\begin{eqnarray}
P(x,t)&&=\frac{x_0^{\lambda/2\Delta+1/2}}{x^{\lambda/2\Delta-1/2}}
\frac{\exp\left(-\frac{x^2+x_0^2}{2\Delta t}\right)}{\Delta t}
I_{\frac{1}{2}+\frac{\lambda}{2\Delta}} \left(\frac{xx_0}{\Delta
t}\right). \label{exact}
\\
W(t)&&=\frac{2\Delta
x_0^{1+\lambda/\Delta}}{\Gamma(1/2+\lambda/2\Delta)}
\exp(-x_0^2/2\Delta t)(2\Delta
t)^{-\frac{3}{2}-\frac{\lambda}{2\Delta}}~. \label{exactabs}
\end{eqnarray}

In the present case for $\lambda\neq 0$ and $\gamma\neq 0$ the
problem corresponds to the motion of a particle in a harmonic
potential with a centrifugal barrier at $x=0$. The spectrum is
discrete and becomes continuous for $\gamma=0$. The problem is
readily analyzed in terms of confluent hypergeometric functions,
more specifically Laguerre polynomials
\cite{Lebedev72,Gradshteyn65}. Incorporating the initial condition
$P(x,0)=\delta(x-x_0)$ and introducing the time scaled variables
\begin{eqnarray}
&&x=\tilde{x}\exp(-\gamma t/2),
\\
&&x_0=\tilde{x}_0\exp(+\gamma t/2),
\end{eqnarray}
we find for $P(x,t)$
\begin{eqnarray}
P(x,t)=\frac{\tilde{x}_0^{\lambda/2\Delta+1/2}}
{\tilde{x}^{\lambda/2\Delta-1/2}} \frac{\gamma e^{\gamma
t/2}}{\Delta\sinh{\gamma t}}
\exp\left[-\frac{\gamma(\tilde{x}^2+\tilde{x}_0^2)}{2\Delta\sinh{\gamma
t}}\right] I_{\frac{1}{2}+\frac{\lambda}{2\Delta}}
\left(\frac{\gamma}{\Delta}\frac{\tilde{x}\tilde{x}_0}{\sinh\gamma
t}\right), \label{exact2}
\end{eqnarray}
and correspondingly for the absorbing state distribution
\begin{eqnarray}
W(t)=\frac{2\Delta
\tilde{x}_0^{1+\lambda/\Delta}}{\Gamma(1/2+\lambda/2\Delta)}
\exp\left[{-\frac{\gamma\tilde{x}_0^2}{2\Delta\sinh{\gamma
t}}}\right]\exp(\gamma t)\left(\frac{\gamma}{2\Delta\sinh\gamma
t}\right)^{\frac{3}{2}+\frac{\lambda}{2\Delta}}. \label{first}
\end{eqnarray}
In Eqs. (\ref{exact}) and (\ref{exact2}) $I_\nu$ is the Bessel
function of imaginary argument, $I_\nu(z)=(-i)^\nu J_\nu(iz)$
\cite{Mathews73}.

\section{\label{secsum}Discussion and conclusion}
Focussing on the expression (\ref{first}) for the
first-passage-time distribution $W(t)$ we note that the damping
constant $\gamma$ defines two distinct time regimes, $1/\gamma$
setting the characteristic cross-over time. In the long time limit
for $\gamma t\gg 1$ the damping constant controls the behavior of
$W(t)$. From Eq. (\ref{first}) we infer
\begin{eqnarray}
W(t)\propto \frac {2\Delta
x_0^{1+\lambda/\Delta}}{\Gamma(1/2+\lambda/2\Delta)}
\left(\frac{\gamma}{\Delta}\right)^{3/2+\lambda/2\Delta}
\exp[-\gamma(1+\lambda/\Delta)t],
\end{eqnarray}
i.e., $W(t)$ falls off exponentially with an effective damping
constant $\gamma(1+\lambda/\Delta)$ renormalized by the ratio
$\lambda/\Delta$ of the nonlinear strength to the noise strength.
We note that for $\Delta\rightarrow 0$ the result is in accordance
with the weak noise phase space derivation in Sec. \ref{secwkb}.
In the intermediate time regime for $\gamma t\ll 1$ the damping
constant $\gamma$ drops out and we obtain
\begin{eqnarray}
W(t)\propto \frac {2\Delta
x_0^{1+\lambda/\Delta}}{\Gamma(1/2+\lambda/2\Delta)}
\exp(-x_0^2/2\Delta t)\left(\frac{1}{2\Delta
t}\right)^{3/2+\lambda/2\Delta}.
\end{eqnarray}
For $2\Delta t\gg x_0^2$ the distribution $W(t)$ exhibits a power
law behavior with the same exponent $3/2+\lambda/2\Delta$ as in
the undamped case for $\gamma=0$. For weak noise this result is
again in agreement with the estimate in Sec. \ref{secwkb}.

\begin{figure}
\includegraphics[width=.8\hsize]{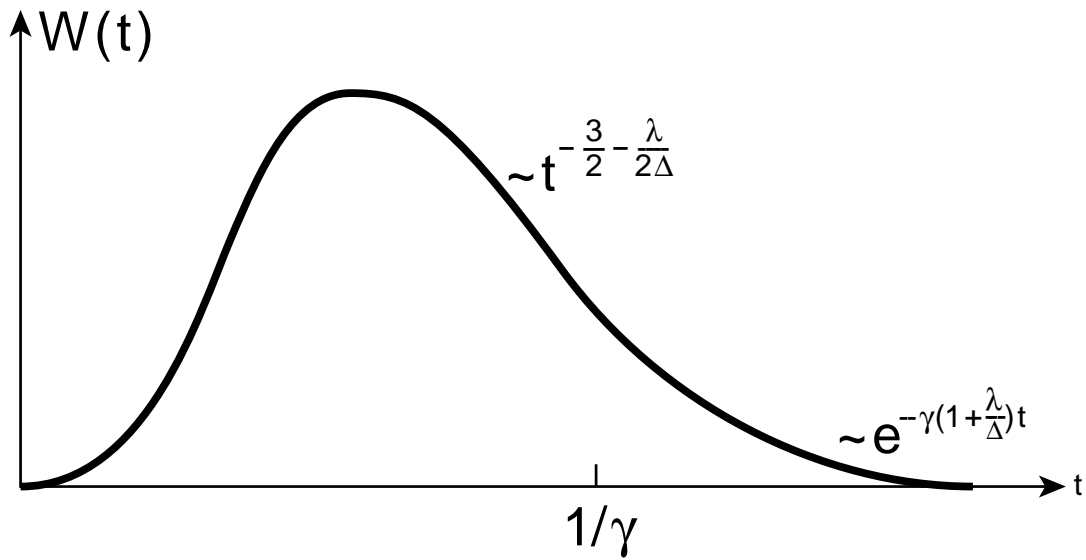}
\caption{We sketch the first-passage-time distribution $W(t)$ as a
function of $t$. In the limit $\rightarrow t 0$ $W(t)$ vanishes
exponentially; about the finite-time-singularity $W(t)$ exhibits a
maximum. At intermediate times for $\gamma t\ll 1$ the
distribution exhibits a power law behavior with scaling exponent
$3/2+\lambda/2\Delta$. In the long time limit for $\gamma t\gg 1$
an exponential fall-off with time constant
$\gamma(1+\lambda/\Delta)$ characterizes the behavior of $W(t)$}
\label{fig3}
\end{figure}

\begin{figure}
\includegraphics[width=.8\hsize]{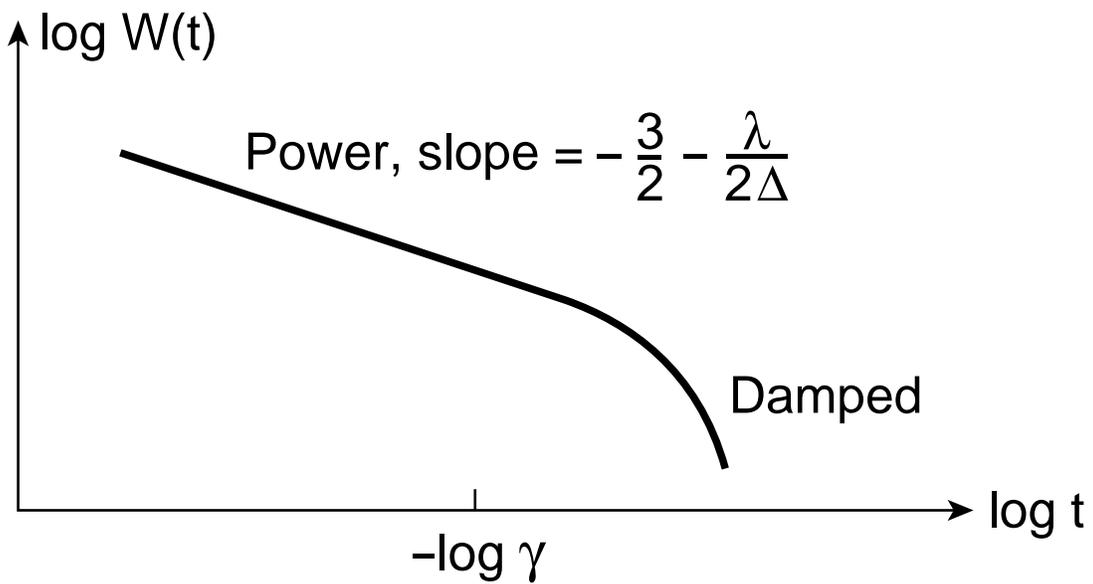}
\caption{In this figure we sketch the behavior of $W(t)$ in a
log-log plot. At intermediate times earlier that $1/\gamma$ we
have scaling behavior with exponent $3/2+\lambda/2\Delta$,
corresponding to a constant negative slope. In the long time limit
the curve dips down indicating the cross-over to exponential
behavior.} \label{fig4}
\end{figure}

In the short time limit $W(t)$ vanishes exponentially and shows a
maximum about the finite-time-singularity. In Fig.~\ref{fig3} we
have depicted the first-passage-time-distribution as a function of
t. In Fig.~\ref{fig4} we illustrate the behavior of $W(t)$ in a
log-log representation.

In this paper we have extended the model discussed in ref.
\cite{Fogedby02d} to include a linear damping term. Not
surprisingly, the damping changes the long time behavior of the
physically relevant first-passage-time distribution. The
finite-time-singularity occurring at time $t_0$ in the noiseless
case is still effectively resolved by the noise, becoming a random
event, but the power law scaling behavior with scaling exponent
$\alpha=3/2+\lambda/2\Delta$ is limited to early times compared
with the cross-over time $1/\gamma$ set by the damping constant.
In the long time limit beyond $1/\gamma$ the damping gives rise to
an exponential fall-off and the scaling property ceases to be
valid. To the extent that the present simple model might apply to
physical phenomena where damping is always present, we must
conclude that an eventual power law scaling presumably is confined
to a time window determined by the size of the damping.
\begin{acknowledgments}
Discussions with A. Svane are gratefully acknowledged.
\end{acknowledgments}
\appendix
\section{\label{app1}Exact solution of the Fokker-Planck equation}
In this appendix we discuss the exact solution of the
Fokker-Planck equation in more detail. Denoting the normalized
eigenfunction of $H$ in Eq. (\ref{ham2}) and the associated
eigenvalues by $\Psi_n$ and $\Delta^2E_n/2$, respectively, we
obtain, incorporating the initial condition $P(x,0)=\delta(x-x_0)$
and the gauge transformation, the following expression for the
distribution
\begin{eqnarray}
P(x,t) = \sum_n e^{-\Delta E_nt/2}e^{-\gamma(x^2-x_0^2)/2\Delta}
(x/x_0)^{-\lambda/2\Delta}\Psi_n(x)\Psi_n^{\ast}(x_0).
\label{prob}
\end{eqnarray}
By means of the transformation $\Psi(x)=x^{1+\lambda/2\Delta}
\exp{-\gamma x^2/2\Delta}G(\gamma x^2/\Delta)$ it follows that $G$
is a solution of the degenerate hypergeometric equation
\cite{Lebedev72,Gradshteyn65}. For the discrete spectrum we choose
the polynomial form and further analysis shows that the
eigenfunctions $\Psi_n$ are given in terms of the Laguerre
polynomials $L_n^\alpha$ \cite{Lebedev72,Gradshteyn65}. For the
normalized eigenfunctions we thus obtain
\begin{eqnarray}
\Psi_n =
\left[2\left(\frac{\gamma}{\Delta}\right)^{3/2+\lambda/2\Delta}
\frac{\Gamma(n+1)}{\Gamma(3/2+\lambda/2\Delta+n)}\right]^{1/2}
x^{1+\lambda/2\Delta}e^{-\gamma x^2/2\Delta}
L_n^{1/2+\lambda/2\Delta}(\gamma x^2/\Delta),
\label{eigfun}
\end{eqnarray}
with discrete eigenvalue spectrum
\begin{eqnarray}
E_n = 4n\gamma/\Delta + 2(\gamma/\Delta)(1+\lambda/\Delta)~.
\label{eigval}
\end{eqnarray}
Inserting Eqs. (\ref{eigfun}) and (\ref{eigval}) in Eq.
(\ref{prob}) and using the identity \cite{Lebedev72,Gradshteyn65}
\begin{eqnarray}
\sum_{n=0}^\infty\frac{n!}{\Gamma(n+\alpha+1)}z^n
L^\alpha_n(x)L^\alpha_n(y) =
\frac{(xyz)^{-\alpha/2}}{1-z}e^{-z(x+y)(1-z)}I_\alpha(2(xyz)^{1/2}/(1-z))~,
\end{eqnarray}
we finally obtain Eq. (\ref{exact2}) for $P(x,t)$ and by the same
analysis as in Ref. \cite{Fogedby02d} the expression (\ref{first})
for $W(t)$.

We note that for $\lambda\rightarrow 0$, using $I_{1/2}(x)=(2/\pi
x)^{1/2}\sinh x$ \cite{Lebedev72}, the expression (\ref{exact2})
takes the form
\begin{eqnarray}
P(x,t) = \left[\frac{\gamma}{\pi\Delta(1-e^{-2\gamma
t})}\right]^{1/2} \left[\exp\left(-\frac{\gamma}{\Delta}
\frac{(x-x_0 e^{-\gamma t})^2}{1-e^{-2\gamma t}}\right)-
\exp\left(-\frac{\gamma}{\Delta} \frac{(x+x_0 e^{-\gamma
t})^2}{1-e^{-2\gamma t}}\right)\right].~~
\end{eqnarray}
i.e., the mirror case of the noise driven overdamped oscillator.
Note 1) that for $\Delta\rightarrow 0$ the variable $x$ lies on
the noiseless orbit $x\rightarrow x_0 \exp(-\gamma t)$ and 2) $P$
vanishes for $x=0$.
\section{\label{app2}Small noise limit - saddle point analysis}
In this appendix we perform for completion a weak noise saddle
point analysis of the exact expression in Eq.(\ref{exact2}) along
the same lines as in ref. \cite{Fogedby02d}. This analysis
requires that we consider both large order and argument of the
Bessel function $I_\nu(x)$. This is easily done by Laplace's
method using a convenient spectral representation
\cite{Lebedev72,Gradshteyn65}
\begin{eqnarray}
I_{\nu}(z)=\frac{(z/2)^\nu}{\Gamma(\nu+1/2)\Gamma(1/2)}
\int_0^\pi\cosh(x\cos\theta)\sin^{2\nu}\theta~d\theta.
\label{spec}
\end{eqnarray}
Inserting (\ref{spec}) in (\ref{exact2}) we have
\begin{eqnarray}
P(x,t) =&& \frac{1}{4\pi\sqrt 2}\frac{1}{\tilde{x}_0}
\frac{\lambda}{\Delta}e^{\lambda/2\Delta}
\left(\frac{\gamma\tilde{x}_0^2}{\lambda\sinh\gamma t}\right)^
{1/2+\lambda/2\Delta}
e^{-(\lambda/4\Delta)(\tilde{x}^2+\tilde{x}_0^2)/
(\tilde{x}\tilde{x}_0\sinh u)} e^{\gamma t/2}\times \nonumber
\\
&&\int_0^\pi d\theta~\frac{\sin\theta}{\sinh u} \left(
e^{(\lambda/\Delta)(\log\sin\theta+(1/2)\cos\theta/\sinh u)} +
e^{(\lambda/\Delta)(\log\sin\theta-(1/2)\cos\theta/\sinh
u)}\right),~~~~
\end{eqnarray}
where $u$ is defined by $\sinh u = \lambda\sinh\gamma
t/(2\gamma\tilde{x}\tilde{x}_0)$, $\tilde{x}=x\exp(\gamma t/2)$,
and $\tilde{x}_0=x_0\exp(-\gamma t/2)$. Setting
$f_\pm(\theta)=\log\sin\theta\pm(1/2)\cos\theta/\sinh u$ the
saddle points for small $\Delta$ are given by
$\cos\theta_\pm=\pm\exp(-u)$ for $x>0$ and
$\cos\theta_\pm=\mp\exp(u)$ for $x<0$. For $x>0$ we have
$f_+(\theta_+)=f_-(\theta_-)=(1/2)(\log(1-e^{-2u})+e^u/\sinh u)$
and $f''_+(\theta_+)=f''_-(\theta_-)=-\coth u$, and we obtain the
weak noise result for $x>0$
\begin{eqnarray}
P(x,t)= \left(\frac{\lambda}{4\pi\Delta}\right)^{1/2}
\frac{e^{\gamma t/2}}{\tilde{x}_0} \left(
\frac{\gamma\tilde{x}_0^2(1-e^{-2u})} {\lambda\sinh\gamma
t}\right)^{1/2+\lambda/2\Delta} \frac
{e^{-(\lambda/2\Delta)(\tilde{x}^2+\tilde{x}_0^2-
2\tilde{x}\tilde{x}_0\cosh u)/\tilde{x}\tilde{x}_0\sinh u}}
{(\sinh u\cosh u)^{1/2}}.~~ \label{disfin}
\end{eqnarray}
For $\Delta\rightarrow 0$ the factor
$\tilde{x}^2+\tilde{x}_0^2-2\tilde{x}\tilde{x}_0\cosh u$ in the
exponent in (\ref{disfin}) locks onto zero, thus setting $
\tilde{x}^2+\tilde{x}_0^2-2\tilde{x}\tilde{x}_0\cosh u = 0 $ and
inserting $\sinh u=\lambda\sinh(\gamma
t)/2\gamma\tilde{x}\tilde{x}_0$ we obtain
\begin{eqnarray}
\tilde{x}^2+\tilde{x}_0^2-((2\tilde{x}\tilde{x}_0)^2+
(\lambda\sinh\gamma t/\gamma)^2)^{1/2} =0.
\end{eqnarray}
Finally, setting $\tilde{x}=x\exp(\gamma t/2)$ and
$\tilde{x}_0=x_0\exp(-\gamma t/2)$ we obtain after some reduction
\begin{eqnarray}
x(t)=\sqrt{x_0^2\exp(-2\gamma t) -
(\lambda/2\gamma)(1-\exp(-2\gamma t))},
\end{eqnarray}
which by simple inspection is equivalent to to Eq. (\ref{sol}. For
$\gamma\rightarrow 0$ we have $x=\sqrt{ x_0^2-\lambda t}$; for
$\lambda\rightarrow 0$, $x=x_0 e^{-\gamma t}$.

\newpage
\end{document}